# Flow and Noise Control of a Cylinder with Annular Plates

Faezeh Eydi

Department of Mechanical Engineering, K. N. Toosi University of Technology, Tehran, Iran
* Corresponding author: Faezeh Eydi (Email: eydi1377@email.kntu.ac.ir)

## Abstract

With the growing demand for electricity and increasing interest in renewable energy, wind turbine noise pollution has become a more pressing issue. Aerodynamic noise not only impacts human well-being but also affects wildlife. This study investigates the aeroacoustics and aerodynamics of a two-dimensional circular cylinder exposed to turbulent flow at a fixed Reynolds number of 22,000. A key novelty of this research is the introduction of a passive control technique, achieved by designing a structure with two thin annular plates positioned around the cylinder in the circumferential space. The numerical analysis is conducted using unsteady Reynolds-averaged Navier-Stokes (URANS) methods in combination with the Ffowcs Williams and Hawkings (FW-H) acoustic analogy. The control assessments reveal substantial performance improvements, with reductions of 89.46% in the mean drag coefficient and 86.6% in the Strouhal number for the modified cylinder. From an acoustic perspective, the passive control method achieves a maximum noise reduction of 39 dB, with a notable drag reduction of 50.62% in a specific case. This innovative design offers a highly efficient and cost-effective solution for noise reduction and wake control in wind turbine applications. Its eco-friendly nature, ease of implementation, and maintenance-free operation make it a promising approach for enhancing wind turbine performance without additional energy consumption.

# 1. Introduction

With the rising global demand for electricity, wind turbines have experienced significant development due to their numerous benefits. However, despite their advantages, they contribute to environmental noise pollution. Wind turbine noise (WTN) primarily originates from mechanical and aerodynamic sources, with the latter being the dominant contributor [1]. The high levels of aerodynamic noise (AN) generated by wind turbines negatively impact human well-being, and pose risks to wildlife [2].

The primary health effects of turbine noise include annoyance, reduced concentration, and sleep disturbances [3]. Individuals with high noise sensitivity are particularly vulnerable, exhibiting stronger emotional responses to noise exposure [4]. Regardless of the noise source, chronic exposure acts as a psychosocial stressor, potentially leading to maladaptive psychological reactions and adverse health effects by influencing the autonomic nervous, neuroendocrine, and immune systems [5]. Studies have linked long-term exposure to community noise with various health issues, including hearing impairment, vertigo, anti-social behavior, blood pressure [6], cardiovascular diseases [7], and increased cortisol levels [8].

In addition to broadband noise, vortex-induced vibration (VIV) is a critical consequence of vortex shedding in cylindrical structures [9, 10]. Despite their simple geometry, circular cylinders are widely used in engineering applications such as wind turbines, towers, high-rise buildings, offshore structures, industrial chimneys, and suspension bridge cables. VIV can lead to structural fatigue failure, making wake control essential for ensuring the longevity of these structures [11, 12]. Flow control techniques for noise and drag reduction are generally classified into active and passive methods. Active control involves external energy sources, such as, such as blowing and suction [13, 14], thermal effects [15-18], plasma actuators [19-21], electromagnetic force [22, 23], feedback control strategies [24-27], rotation [28, 29], rotating control rods [30-32], and synthetic jets [33-36]. Conversely, passive control focuses on modifying the structure's surface shape to enhance aerodynamic performance. Examples include slotted circular cylinder [37-39], porous media [40-42], helical strakes [43-46], control rods [47-49], rigid and flexible splitter plates [50-56], shape modification [57-61] and surface bumps [62, 63]. A key advantage of passive methods is their cost-effectiveness, as they require no additional energy input and minimal maintenance [64].

Extensive research has been conducted on noise reduction in circular cylinders. Eydi and Mojra [65] numerically investigated the effects of two arc plates on flow and noise control around a cylinder in turbulent flow. Their results showed that incorporating arc plates significantly reduced both noise levels and the drag coefficient, achieving a maximum noise reduction of 27 dB and a 39% decrease in drag. In another research, Eydi et al. [66] examined the influence of three arc plates combined with flexible splitter plates in the single and dual forms on flow control. Their findings demonstrated that the drag could be reduced by 50% with the addition of circular arc plates. Xu et al. [67] conducted an experimental investigation aimed at minimizing flow drag and noise generated by a circular cylinder utilizing a porous material plate positioned downstream. The findings indicated that the plate significantly reduces both pressure drag and noise levels, demonstrating a notable advantage over the impermeable splitter plate and porous coating. Moradi and Mojra [68] explored the use of porous-filled grooves for noise reduction and turbulence control through a parametric study that assessed groove number, arrangement, size, and permeability. Their results indicated that positioning grooves at 90° from the front stagnation point reduced the overall sound pressure level (OASPL) by 1.25 dB. Similarly, Ansari et al. [40] investigated porous coatings for flow noise reduction, reporting that the drag coefficient reduced by over 50% compared to a bare cylinder. Additionally, their study recorded a maximum noise reduction of 6 dB across. Fujisawa et al. [69] experimentally examined the aerodynamic noise reduction of a circular cylinder using longitudinal grooves. Their results revealed that when the groove depth-to-diameter ratio was set to 0.010 and 0.017, noise levels decreased by 8 dB and 5 dB, respectively. Maryami et al. [70] studied the effects of symmetric local blowing at different angles on noise suppression. Their findings showed that local blowing delayed vortex shedding, extended shear layers, and shifted the high-momentum transfer region further downstream, leading to noise reduction. Al-Sadawi et al. [71] investigated the use of plasma actuators on a flat plate with a blunt trailing edge, demonstrating that plasma-induced jet magnitudes of 9–10% and 7% of the freestream velocity (for downward and spanwise plasma actuators, respectively) were sufficient to suppress vortex shedding tonal noise.

Given the increasing demand for renewable energy and the adverse effects of fossil fuel dependency, restricting wind energy exploitation is not economically viable. Wind turbines offer numerous advantages, including clean energy production, reduced water consumption, and high power output. However, their noise impact must be addressed through eco-friendly designs and noise reduction strategies. The significance of noise pollution on human health, wildlife, and overall quality of life has motivated further research in this field. This study introduces the use of annular plates as a passive flow control method for noise reduction. A comprehensive parametric analysis is conducted, focusing on the normalized annular diameter and the open angle between the annular plates to optimize noise and drag reduction performance. These plates offer several advantages, including noise reduction, mitigation of fatigue damage, extended structural lifespan, ease of installation, and cost-effectiveness. The study employs a two-dimensional unsteady Reynolds-Averaged Navier-Stokes (URANS) solver and the Ffowcs Williams and Hawkings (FW-H) acoustic analogy to ensure reliable results.

The remainder of this paper is structured as follows: Section 2 presents the numerical methodology, grid independence study, and simulation model validation. Section 3 discusses the results, and Section 4 concludes the study.

## 2. Material and methods

This section introduces a novel configuration of a circular cylinder featuring circumferential annular plates, aimed at enhancing noise reduction and optimizing flow control. To this end, Section 2.1 presents the problem description, while Section 2.2 details the numerical methodology. Section 2.3 introduces the non-dimensional parameters relevant to the study. A grid independence study, essential for optimal grid design, is discussed in Section 2.4. Finally, to validate the numerical model, the results are compared with previous experimental studies in Section 2.5.

### 2.1. Problem description

In the present study, a configuration consisting of two thin annular plates positioned around a cylinder is proposed for flow and noise control (Fig. 1). These plates are placed symmetrically and concentrically at the top and bottom of the cylinder. The arrangement ensures that the front and rear sections of the cylinder remain open, allowing the fluid to flow through the space between the solid boundaries. The annular diameter and the coverage angle of the open arc between the annular plates are denoted as $D_a$ and $\theta$, respectively. The diameter of the annular plates is normalized by the cylinder diameter, represented as $D^*$, which varies within the range $1.125 \leq D^* \leq 1.625$. Additionally, the coverage angle $\theta$ ranges from 30° to 120°. The thickness of the annular plates is set to $0.0125D$.

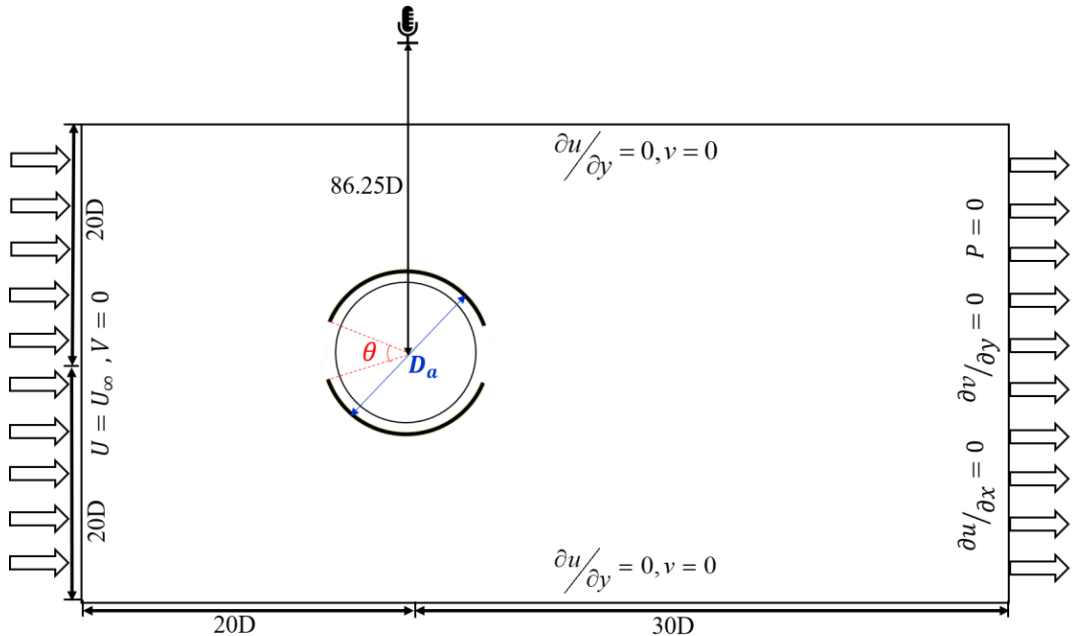

**Fig. 1.** Computational domain and boundary condition with definition of symbols, $\theta$: the open-angle between annular plates, $D_a$: the diameter of annular plates.

The computational domain and boundary conditions, along with the definitions of relevant symbols, are illustrated in Fig. 1. A rectangular computational domain with dimensions 50D×40D is considered, where D represents the cylinder diameter [65]. Given the practical significance of higher Reynolds numbers in engineering applications, the Reynolds number is set to 22,000, corresponding to a Mach number of 0.0058. A freestream velocity is applied at the inlet boundary, while a zero-pressure gradient is imposed at the outlet. A no-slip boundary condition is enforced on the upper and lower boundaries, which are located 20D away from the cylinder center, resulting in a blockage ratio of approximately 2.5%.

## 2.2. Governing equations

All aeroacoustics parameters can typically be computed using the Navier-Stokes equations. However, extracting the nonlinear terms related to noise generation and propagation from these equations is a time-consuming process. To identify all noise sources, the relevant fluid flow parameters must first be determined. Once these parameters such as hydrodynamic forces over time are obtained, the Ffowcs Williams and Hawkings (FW-H) acoustic analogy is used to predict sound generation [65].

### 2.2.1. Fluid flow
The continuity and momentum equations for flow simulation are utilized as follow:

$$\frac{\partial u_i}{\partial x_i} = 0 \tag{1}$$

$$\frac{\partial u_i}{\partial t} = -\frac{1}{\rho}\frac{\partial u}{\partial x_i} + v\frac{\partial^2 u_i}{\partial x_j \partial x_j} - \frac{\partial}{\partial x_i}(u_i u_j - \overline{u'_i u'_j}) \tag{2}$$

Where u is the fluid velocity, $\rho$ is density and p is static pressure. $x_1$ and $x_2$ denote the streamwise and cross-stream direction, respectively. The fluctuation part of velocities and Reynolds stress are defined in $u'_i$, $u'_j$ and $\overline{u'_i u'_j}$, respectively.

### 2.2.2. Acoustic
Aside from the Navier–Stokes equations, the Ffowcs Williams–Hawkins (FW–H) equation can be used to estimate aeroacoustics characteristics. First introduced in 1969 [72], the FW–H equation models the arbitrary motion of sound radiation in the far field. As a widely used and powerful analytical approach [73], it accounts for all noise sources (monopole, dipole, and quadrupole) represented in Eq. (3). This acoustic analogy is particularly valued for its accuracy in noise estimation, benefiting from the high-resolution aerodynamic calculations provided by the k–ω SST turbulence model.

$$\frac{1}{c_0^2}\frac{\partial^2 P'}{\partial t^2} - \nabla^2 P' = \frac{\partial^2}{\partial x_i \partial x_j}[T_{ij}H(f)] - \frac{\partial}{\partial x_i}\left[[P_{ij}n_j + \rho u_i(u_n - v_n)]\delta(f)\right] \\ + \frac{\partial}{\partial t}[[\rho_0 v_n + \rho(u_n - v_n)]\delta(f)] \tag{3}$$

Here, $C_0$ is the sound speed, $H_f$ and $\delta(f)$ are Heaviside and Dirac Delta functions, respectively. $\tau_{ij}$ is viscous stress, $\delta_{ij}$ represents the Kronecker delta and sound pressure of far- fields represent by $P'$. $n_i$ indicates the perpendicular unit vector on the outer surface of planes that exist in the numerical domain. Moreover, we proposed a new parameter called *f*. The magnitude of this parameter for all rigid surfaces, including cylinder and annular plates, is equal to 0 (*f*=0), and for boundless space, this value is positive (*f* > *0*). $T_{ij}$ denotes the Lighthill stress tensor, which is quadrupole, and $P_{ij}$ is the compressive stress tensor for Stokes flow. To compute these two tensors, Eq. (4) and Eq. (5) are provided [65].

$$T_{ij} = \rho u_i u_j - \tau_{ij} + \delta_{ij}((p - p_0) - c_0^2(\rho - \rho_0)) \tag{4}$$

$$P_{ij} = p\delta_{ij} - \mu[\frac{\partial u_i}{\partial x_j} + \frac{\partial u_j}{\partial u_i} - \frac{2}{3}\frac{\partial u_k}{\partial x_k}\delta_{ij}] \tag{5}$$

Eventually, Eq. (6) is used to calculate the acoustic pressure. Here in Eq. (6), T indicates the thickness of the monopole sound source determined by the geometry, and L represents the loading of the dipole sound source caused by force exerted on the fluid as a result of the body's presence. So, the terminology is according to corresponding aerodynamics [74]. Eq. (7) and Eq. (8) are used for T and L, respectively [65].

$$P'(\vec{x},t) = P'_T(\vec{x},t) + P'_L(\vec{x},t) \tag{6}$$

$$P'_T(\vec{x},t) = \frac{1}{4\pi}(\frac{\partial}{\partial t}\int_S \left[\frac{\rho_0 n_i\left(1-\frac{\rho}{\rho_0}\right)v_i + \frac{\rho u_i}{\rho_0}}{(r(1-M))}\right]_{ret} ds) \tag{7}$$

$$P'_L(\vec{x},t) = \frac{1}{4\pi}(-\frac{\partial}{\partial x_i}\int_S \left[\frac{P_{ij}n_i + \rho u_i + (u_{i_n} - u_{j_n})}{(r(1-M))}\right]_{ret} ds) \tag{8}$$

### 2.3. Non-dimensional flow parameters

The parameters of flow characteristic include mean drag coefficient ($\overline{C_d}$), Strouhal number ($St$), and mean pressure coefficient ($\overline{C_P}$) are given by [54]:

$$\overline{C_d} = \frac{1}{N}\sum_{i=1}^{N}\frac{F_d}{0.5\rho U_\infty^2 D} \tag{9}$$

$$St = \frac{f_s D}{U_\infty} \tag{10}$$

$$\overline{C_P} = \frac{1}{N}\sum_{i=1}^{N}\frac{P_i - P_\infty}{0.5\rho U_\infty^2} \tag{11}$$

In these relations, $\rho$ is density fluid, $F_D$ is drag force, $f_s$ is the vortex shedding frequency and P and $P_\infty$ are static pressure and reference pressure of inlet boundary, respectively. Here, N is number of samples and $P_i$ denotes the temporal series.

The parameters of acoustic analysis include root mean square of static pressure ($P_{rms}$) and sound pressure level (SPL) are defined below [65]:

$$P_{rms} = \sqrt{\frac{1}{N}\sum_{i=1}^{N}(P_i - \bar{P})^2} \quad (12)$$

$$SPL(dB) = 20 \log \left(\frac{P_{rms}}{P_{ref}}\right) \quad (13)$$

Here, the reference pressure of acoustic is defined by $P_{ref}$, which is equal to $20\mu Pa$.

## 2.4. Mesh topology

As shown in Fig. 2a, a structured mesh is utilized throughout the computational domain to accurately capture extreme gradients and prevent numerical instabilities. Fig. 2b provides a zoomed-in view around the cylinder and annular plates, where the radial grid becomes finer as it approaches the structural surface. The cylinder perimeter is discretized using 360 nodes, ensuring adequate resolution. The cell expansion ratio across the entire mesh is maintained below 1.09. The first layer meshing, placed within the buffer zone, has a thickness of approximately 0.003D, meeting the $y^+_{average}$ of 0.98. Here, $y^+$ stands for the non-dimensional wall distance, which is determined as $y^+ = u_\tau y/v$ ($u_\tau$: friction velocity, $y$: distance from wall and $v$: fluid kinematic viscosity). A grid independence study is conducted for the cylinder with annular plates at $D^* = 1.125$ and $\theta = 30°$ using four different mesh densities. Table 1 confirms the independence of key flow characteristics, including the Strouhal number and the time-averaged drag coefficient. Since further refinement of the fine grid does not significantly impact these parameters, this grid is chosen for the simulations. For the selected mesh, the non-dimensional time step ($\frac{U_\infty \Delta t}{D}$) is set to 0.5, resulting in a Courant number ranging between 1 and 4.5 across the computational domain.

**Table 1.** Grid independency test for the cylinder surrounded by annular plates when $D^* = 1.125$ and $\theta = 30°$.

| Case study | Grid system | Number of elements | $y^+_{average}$ | St | $\overline{C_d}$ |
|---|---|---|---|---|---|
| | Corse | 43600 | 28.989 | 0.238 | 0.637 |
| Cylinder with two annular plates | Medium | 64000 | 13.247 | 0.233 | 0.6301 |
| | Fine | 78800 | 0.98 | 0.23 | 0.628 |
| | Extremely fine | 97000 | 0.576 | 0.23 | 0.628 |

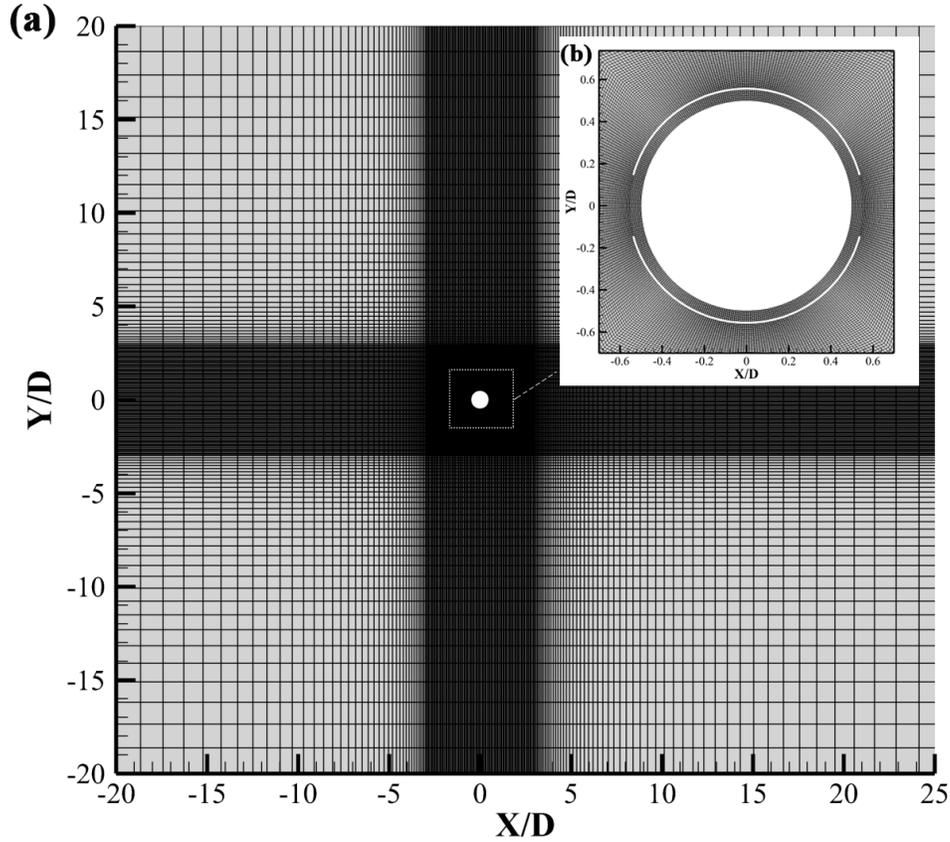

**Fig. 2.** Meshing of the computational domain, (a) view of the whole domain, (b) zoomed view of the mesh around the cylinder and annular plates when $D^* = 1.125$ and $\theta = 30°$.

## 2.5. Validation of simulation model

In this study, a single circular cylinder subjected to a turbulent flow regime is used for validation. The numerical approach is based on the finite element method and implemented using ANSYS Fluent. To accurately capture eddy viscosity, the k-w SST turbulence model is employed. This model is widely used in industrial applications due to its stability and reliable accuracy across a broad range of turbulent flows [75-77]. Table 2 compares key aerodynamic parameters with previous numerical studies, showing good agreement in the Strouhal number and mean drag coefficient. However, the root-mean square (RMS) lift coefficient is slightly higher than in other studies due to limitations in accurately predicting the separation point and inherent shortcomings of two-dimensional approaches. Furthermore, Fig. 3 presents the time-averaged pressure coefficient and streamwise velocity at a wake-region distance equal to the cylinder diameter, compared against experimental data [78]. Both comparisons exhibit strong agreement, confirming the reliability of the mesh resolution for further analysis. To validate the acoustic analogy for noise prediction, the sound pressure level (SPL) is compared with the experimental results of Casalino and Jacob [79] in Fig. 4. As mentioned earlier, noise estimation in this two-dimensional problem is conducted using the k-w SST turbulence model in conjunction with the Ffowcs Williams and Hawkings (FW–H) acoustic analogy [80-82]. The cylinder is treated as a noise source, while a microphone is positioned vertically at 86.25D above the cylinder as an acoustic receiver (Fig. 1) [79]. The numerical Strouhal number is 0.226, whereas the experimental value is 0.2. According to Fig. 4, there is a slight discrepancy (~4 dB) between numerical and experimental SPL results. This difference arises due to the absence of spanwise effects, as well as the limitations of the unsteady Reynolds-averaged Navier-Stokes (URANS) method. The URANS approach assumes acoustic compactness, which is unable to fully resolve small-scale eddies in the near wake, leading to some deviation in SPL predictions. Additionally, the influence of three-dimensional flow structures at high Reynolds numbers contributes to the discrepancy. Despite the inherent limitations of two-dimensional simulations, they effectively capture the essential characteristics of vortex shedding while significantly reducing computational costs. The numerical methods employed demonstrate reasonable accuracy in predicting aerodynamic noise, making them suitable for further investigations into noise generation mechanisms.

**Table 2.** Verification of characteristic flow of aerodynamics for a circular cylinder

| Numerical study | Turbulence Model | $Re$ | $St$ | $\overline{C_d}$ | $Cl_{rms}$ |
|---|---|---|---|---|---|
| Present study | k-$w$ SST | $2.2 \times 10^4$ | 0.226 | 1.12 | 0.58 |
| Orselli et al. [83] | k-$w$ SST | $9 \times 10^4$ | 0.247 | 0.944 | - |
| Kravchenko and Moin [84] | LES | $3.9 \times 10^3$ | 0.210 | 1.04 | - |
| Feng et al. [85] | LES | $3.9 \times 10^3$ | 0.214 | 1.015 | 0.097 |
| Doolan [86] | $k$-$\varepsilon$ Standard | $2.2 \times 10^4$ | 0.24 | 0.98 | 0.37 |

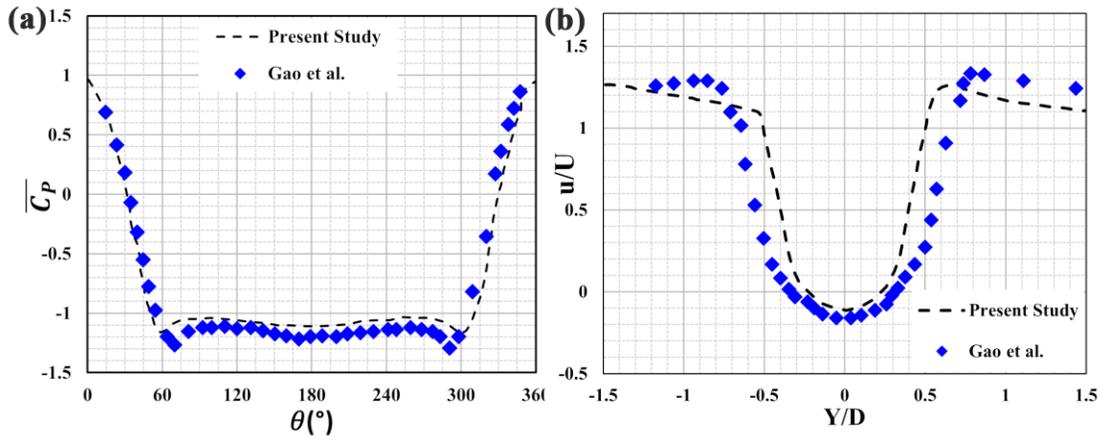

**Fig. 3.** Comparison of the numerical results with experimental [78]: (a) time-average pressure coefficient over the cylinder surface, (b) time-averaged streamwise velocity in the wake region (x/D=1).

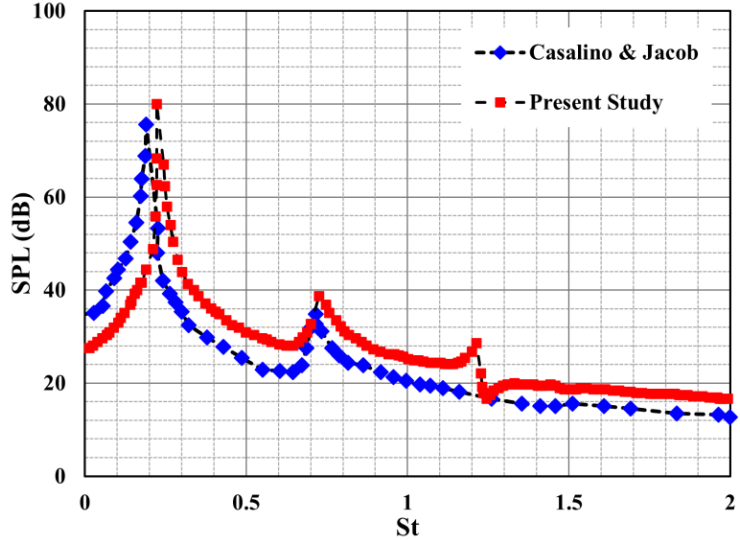

**Fig. 4.** Validation of acoustic analysis with experimental SPL[79].

## 3. Results and discussion

This section presents the results of a two-dimensional circular cylinder incorporating a novel application of URANS methods and the FW–H analogy.

When flow passes over a bluff body, vortices are shed alternately from one side to the other, generating alternating low-pressure zones downstream. This phenomenon, known as vortex shedding, induces oscillations in drag and lift forces, leading to vortex-induced vibrations (VIVs), which can significantly impact the structure's lifespan [54]. Therefore, mitigating VIVs is crucial. In this study, two thin annular plates are employed as a passive flow control method to manipulate dominant flow structures and enhance the fluid dynamics around the circular cylinder. As shown in Fig. 5, this novel passive strategy influences both the Strouhal number and the time-averaged drag coefficient. The results highlight the strong sensitivity of these parameters to the configuration of the annular plates, leading to substantial variations in key vortex shedding characteristics. Due to the closed top and bottom of the cylinder and varying open angles ($\theta$), the oscillation response differs across cases. In most scenarios, equipping the cylinder with annular plates increases the Strouhal number; however, a significant 86.6% reduction is observed for $D^* = 1.25$ when $\theta = 60°$. The most effective mechanism for suppressing oscillations and delaying the separation point is the confinement of the cylinder within the limited space between the structural boundaries of the plates. Conversely, in some cases, particularly at $\theta = 90°$, a sharp increase in the Strouhal number occurs due to flow obstruction by the upper and lower annular plates. Furthermore, the results demonstrate a remarkable capability for drag reduction with this configuration. As shown in Fig. 5b, the drag coefficient decreases across all investigated cases compared to the bare cylinder, even reaching negative values in four cases. The overall trend of this graph is assenting until it reached $\theta = 90°$ and then descended in $\theta = 120°$. The lowest mean drag coefficient occurs at $\theta = 30°$, with the case of $D^* = 1.25, \theta = 30°$ achieving an impressive 89.46% reduction, making it a highly desirable outcome.

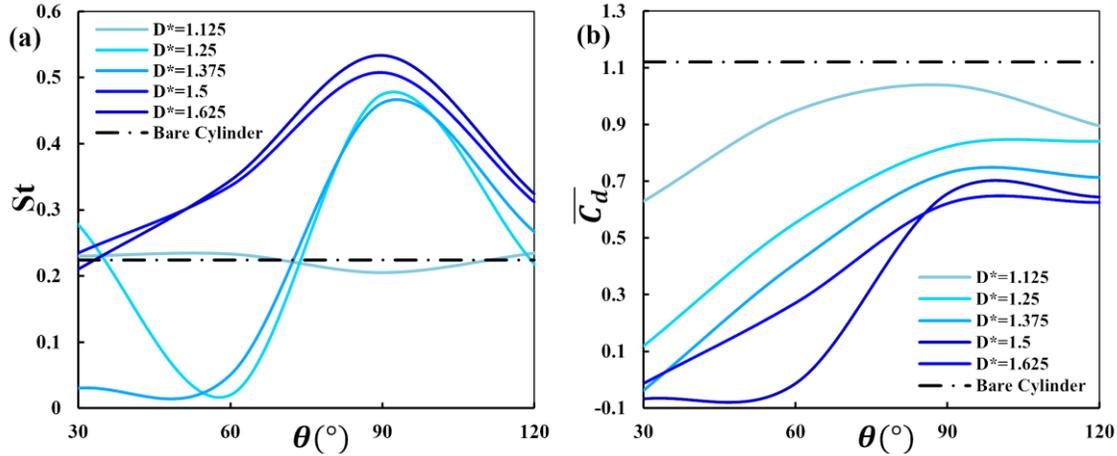

**Fig. 5.** Strouhal number (St) and time-averaged drag coefficient ($\overline{C_d}$) of equipped circular cylinder utilizing annular plates as a passive technique which increases or decreases the Strouhal number. Furthermore, in all cases dramatic reductions of mean drag coefficient occur even, it gets negative in four cases.

The variations in the drag coefficient stem from the pressure drop between the upstream and downstream regions of the cylinder. A smaller pressure difference results in a lower drag coefficient [54]. To further investigate these variations, the time-averaged pressure coefficient contours for all equipped cylinders are presented in Fig. 6. As observed, the introduction of circumferential annular plates significantly alters the pressure distribution. This passive control method forces the flow toward the base point of the cylinder, reducing pressure differences and ultimately leading to noticeable drag reduction. The increase in base pressure relative to the stagnation point pressure plays a crucial role in modulating the mean drag coefficient. For instance, this effect is particularly evident in the case of $D^* = 1.625$, $\theta = 30°$. As previously noted, the maximum and minimum drag reductions across all diameter ratios occur at $\theta = 30°$ and $\theta = 90°$, respectively. To further interpret these findings, a comparison is made between the first and third columns of the mean pressure coefficient distributions. The pressure distribution around the cylinder, especially at the front and rear stagnation points, exhibits distinct variations depending on the plate configuration. For example, in the case of $D^* = 1.5$, the pressure distribution in the space between the cylinder and the plates is relatively high when $\theta = 30°$. However, at $\theta = 90°$, the highest pressure zone shifts to the front stagnation point of the cylinder. These results align well with the trends observed in the mean drag coefficient (Fig. 5b), further validating the impact of annular plates on flow behavior and drag reduction.

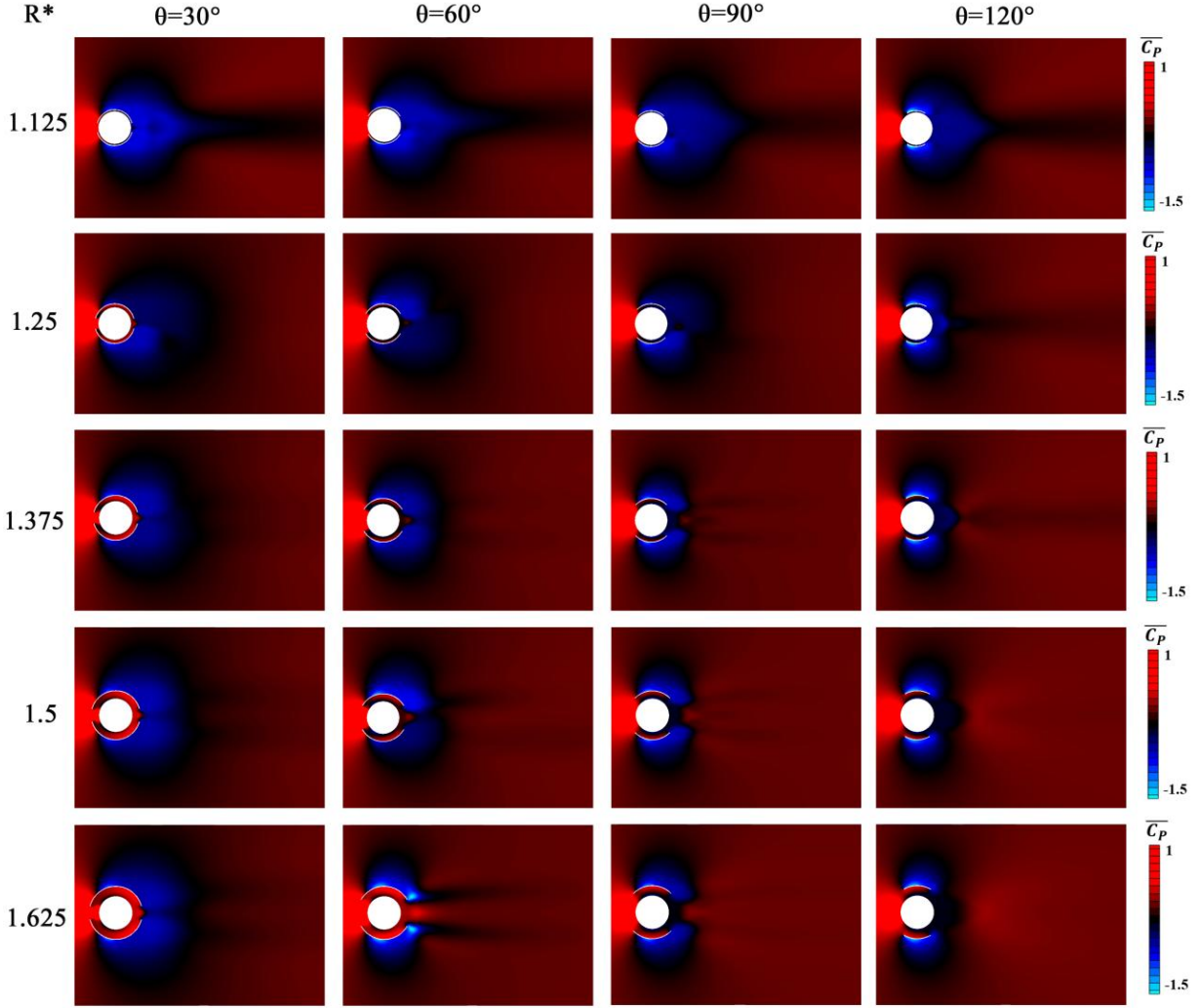

**Fig. 6.** Time-averaged pressure coefficient ($\overline{C_P}$) of equipped cylinders utilizing annular plates as a passive technique which caused creation of a relative pressure balance in cylinder surface and finally reduction of mean drag coefficient.

Instantaneous non-dimensionalized vorticity contours ($W_z^* = \frac{W_z D}{U_\infty}$) for a circular cylinder in a von Kármán vortex street are shown in Fig. 7. These contours illustrate the development of two shear layers, characterized by positive (counterclockwise) and negative (clockwise) vorticity [54]. The shedding process in the unsteady wake varies depending on the configuration. The positions of vortices and the horizontal spacing between successive vortices significantly influence vortex shedding frequency. For instance, the highest Strouhal number (St) is observed for $D^* = 1.625, \theta = 90°$, where a greater number of vortices with shorter spacing is present. This condition indicates increased wake instability and turbulence. Conversely, in some cases, the application of passive control inhibits the interaction between shear layers, effectively suppressing vortex shedding. For example, for an equipped cylinder with $D^* = 1.375$ and $\theta =$ 30°, the shear layer interactions are weak, leading to an independent shedding mechanism for each layer. This behavior contributes to reduced wake instability and a lower shedding frequency.

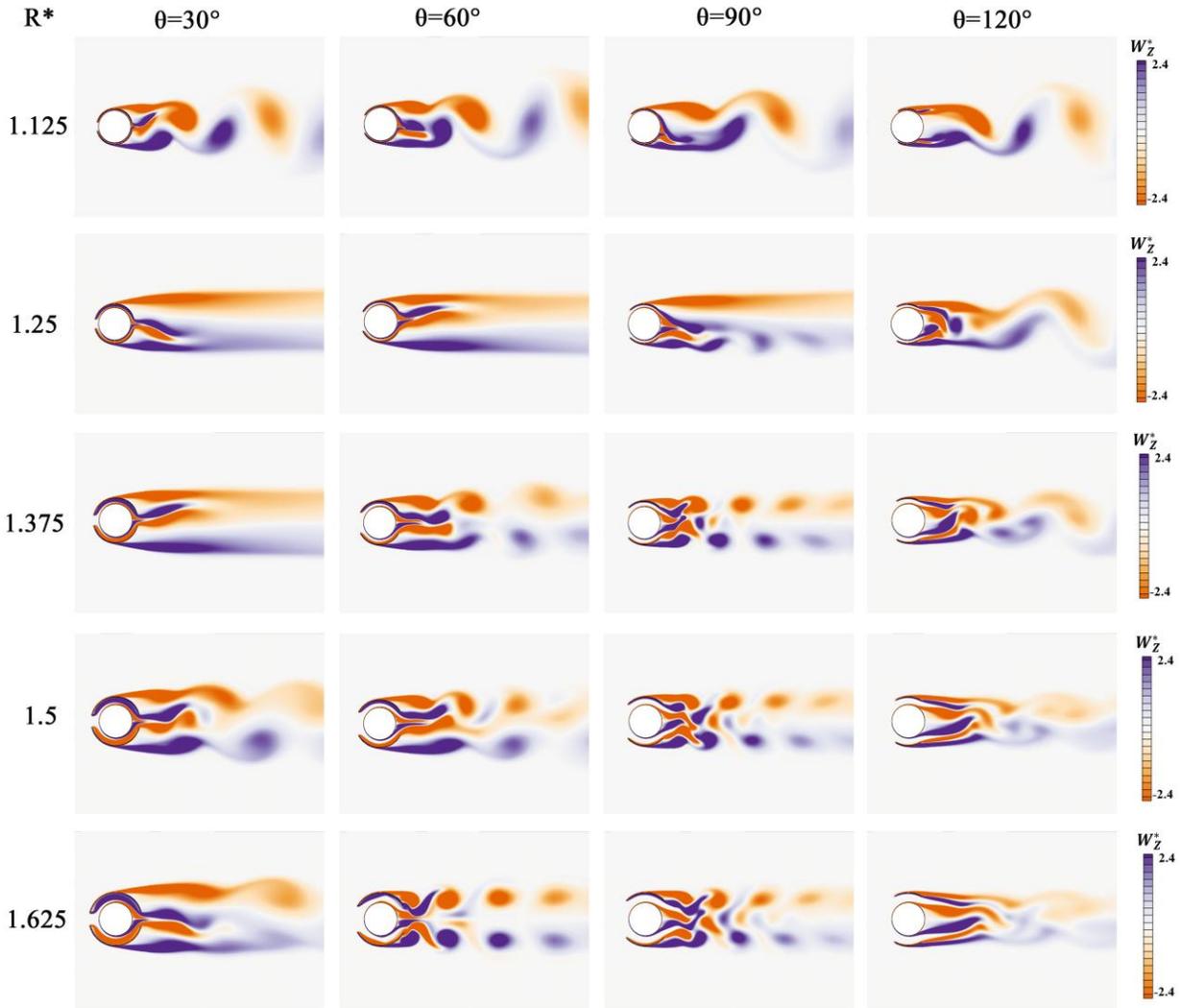

**Fig. 7.** Instantaneous non-dimensionalized contours of vorticity $(W_Z^*)$ for equipped circular cylinders utilizing annular plates as a passive technique which in some cases, vortexes diminish perfectly and in contrast, in other cases, vorticities increased, illustrating more wake destabilization.

Investigating the dynamic mean wake flow in unsteady simulations is essential for understanding vortex shedding behavior. Figures 8 and 9 illustrate the iso-contours of the non-dimensional root mean square (RMS) of the streamwise velocity ($u^* = \frac{u_{rms}}{U_\infty}$) and transverse velocity ($v^* = \frac{v_{rms}}{U_\infty}$) for all case studies. With the implementation of passive flow control, velocity fluctuations are reduced, leading to vortex shedding suppression. The number and intensity of strong production peaks vary depending on the key parameters ($\theta, D^*$), which indicate the roll-up positions of shear layers [87]. In most cases, the strong peaks in $u^*$ appear symmetrically relative to the wake centerline.

The parameter $v_{rms}$ evaluates the interaction between wake vortices and reflects variations in lift force [54]. As lift fluctuations are a dominant source of noise generation [88], $v^*$ serves as a key indicator for assessing the effectiveness of the control approach in reducing aerodynamic noise. Across all cases with equipped cylinders, $v^*$ decreases compared to the bare cylinder, confirming the effectiveness of the passive control method. However, in the case of $D^* = 1.625$, $\theta = 60°$, the sharp rear edges of the annular plates enhance shear layer interactions, resulting in the formation of strong peaks on both sides of the bluff body.

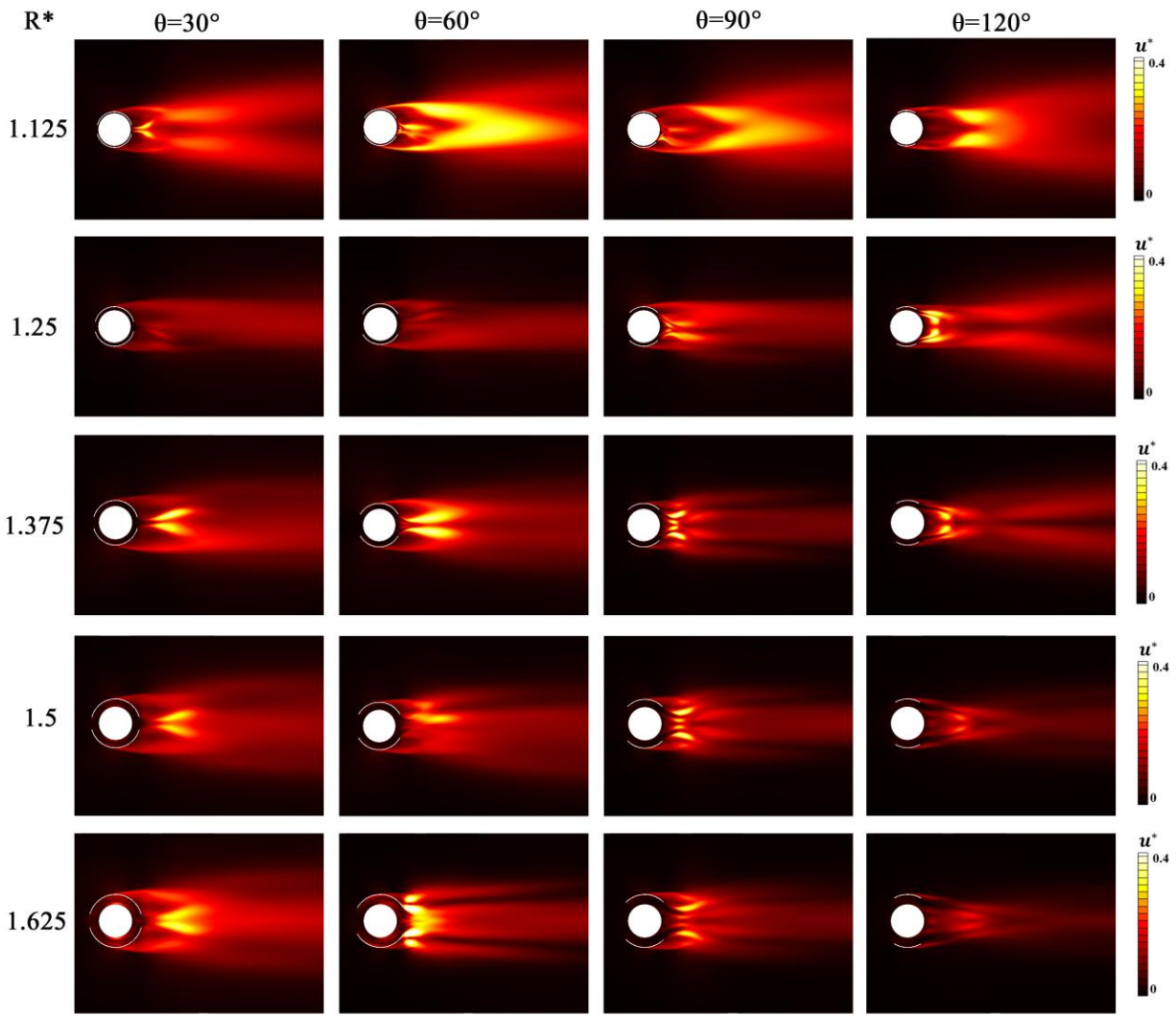

**Fig. 8.** Iso-contours of non-dimensionalized root-mean-square streamwise velocities ($u^* = \frac{u_{rms}}{U_\infty}$) utilizing annular plates as a passive technique which decreased the velocity gradient in the direction perpendicular to flow.

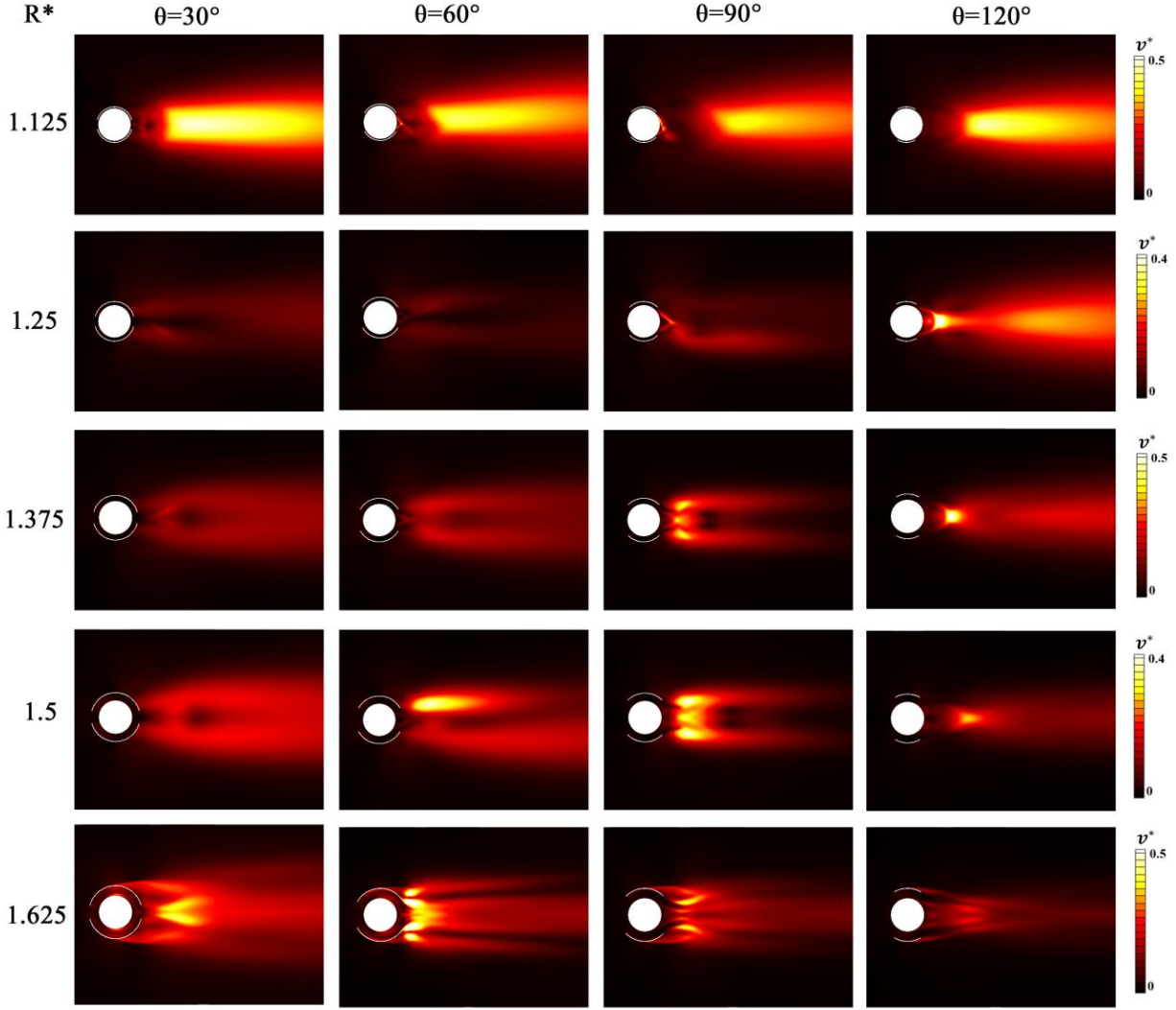

**Fig. 9.** Iso-contours of non-dimensionalized root-mean-square transverse velocities ($v^* = \frac{v_{rms}}{U_\infty}$) utilizing annular plates as a passive technique which decreased the velocity gradient in the direction of flow.

Turbulent mixing in the wake region can be assessed using the turbulent kinetic energy (TKE) [89]. Additionally, TKE serves as an indicator of surface pressure unsteadiness and the fluctuating amplitude of dynamic flow loads acting on the cylinder [90]. As shown in Fig. 10, TKE values are relatively high in the wake region, particularly along vortex formation pathways. Moreover, peak TKE levels occur in wake regions farther from the structural boundary of the cylinder. The presence of annular plates modifies the TKE distribution compared to the bare cylinder, aligning with the trends observed in the time-averaged velocity fluctuations shown in Figs. 8 and 9. Furthermore, a reduction in turbulence intensity is indicative of vortex shedding suppression in three-dimensional flows [69]. From an aeroacoustics perspective, lowering TKE intensity leads to decreased pressure fluctuations, ultimately contributing to noise reduction.

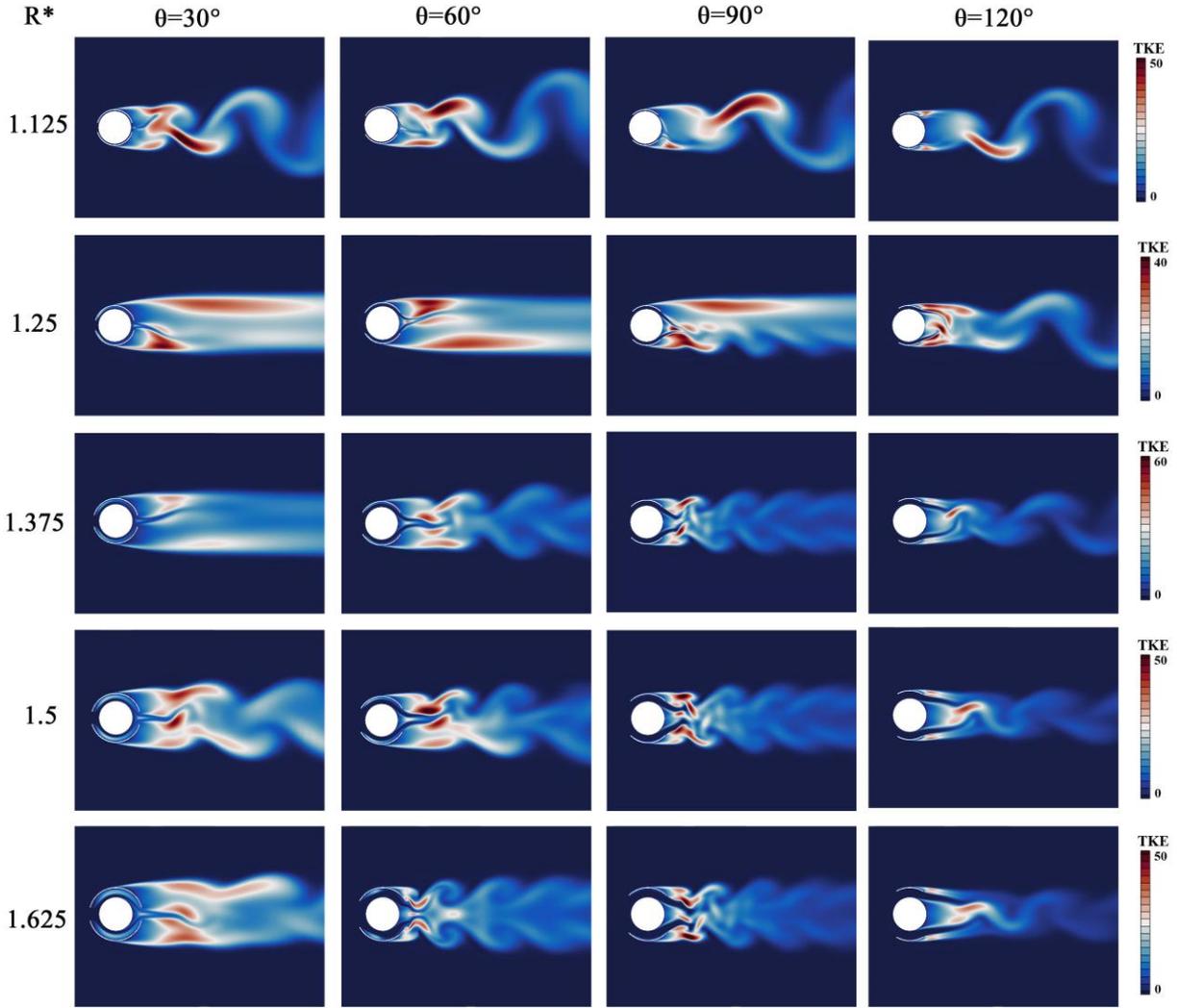

**Fig. 10.** Percentage of turbulence kinetic energy (TKE) around the equipped cylinder utilizing annular cover as a passive control method, which modified the turbulence intensity.

Eddy viscosity, a key characteristic of turbulent flow, is presented in Fig. 11. It acts as a proportionality factor that quantifies the turbulent energy transfer caused by moving eddies, leading to tangential strain. In turbulent flows, momentum transfer is primarily governed by mixing induced by large, energetic turbulent eddies. As previously discussed in the analysis of TKE contours, a similar pattern of modification and forward motion is observed in Fig. 11, indicating wake stabilization. This suggests that the applied passive control method effectively alters the turbulence structure, contributing to improved wake dynamics.

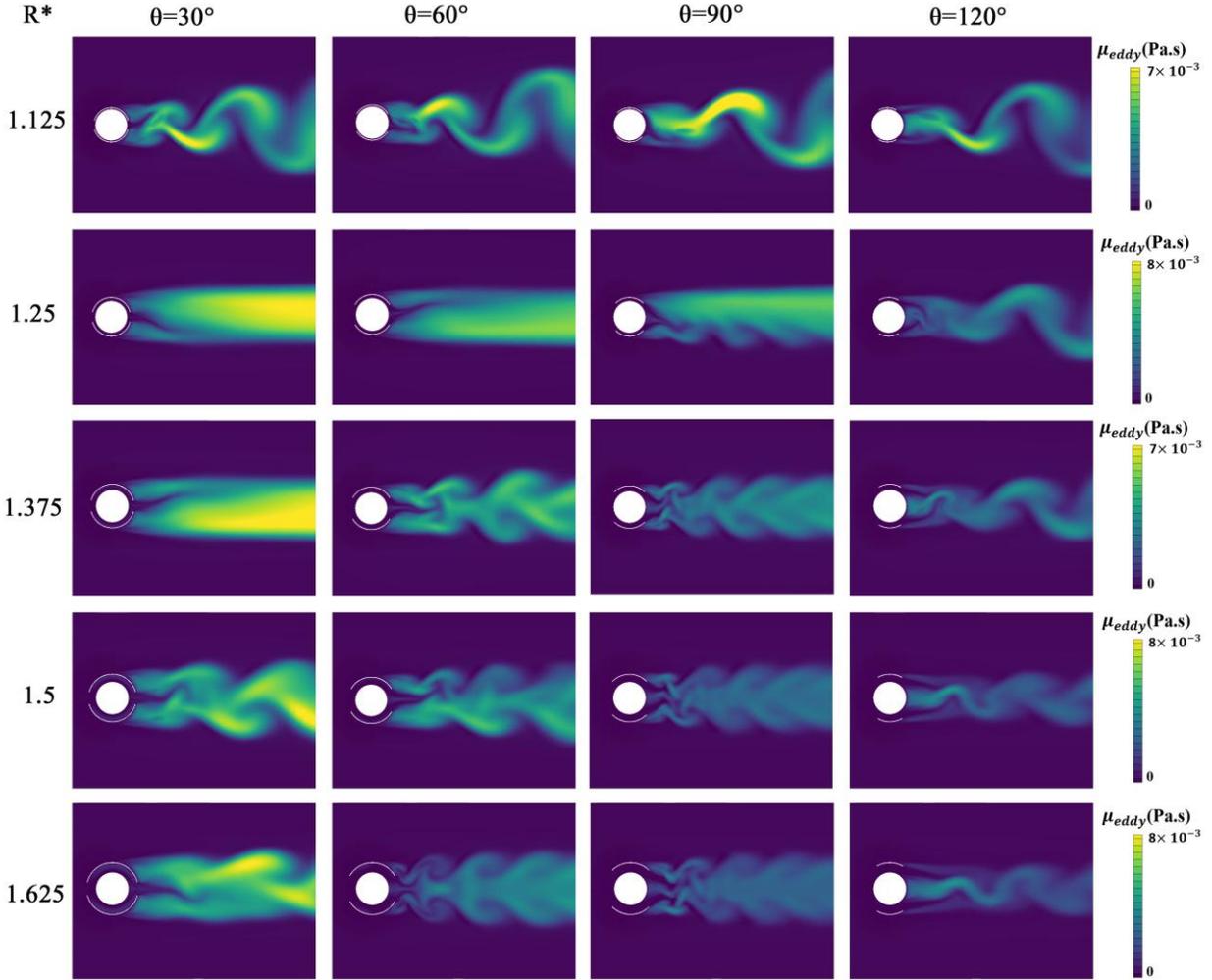

**Fig. 11.** Eddy viscosity around the equipped circular cylinder utilizing annular cover as a passive control method, which has forward motion in wake, illustrating the wake stabilization.

Figure 12 presents the iso-contours of the root mean square of static pressure ($P_{rms}$) for all investigated cases. A high $P_{rms}$ magnitude near solid surfaces indicates intense pressure fluctuations, which play a key role in noise generation and amplitude [65]. The results show that no strong $P_{rms}$ peaks are present around the cylinder surface. In most cases, significant pressure fluctuations occur in the wake region. However, these fluctuations contribute minimally to far-field noise since no lift dipole support exists in areas distant from solid boundaries [65, 91]. Additionally, all equipped cylinders exhibit lower pressure fluctuation levels compared to the bare cylinder, confirming the effectiveness of the passive control method in mitigating noise sources. The highest pressure gradient intensity is observed near the rear structural boundary of the annular plates in the case of $D^* = 1.625$, $\theta = 60°$. Sound pressure level (SPL) is the most critical parameter for evaluating the impact of passive control techniques on aeroacoustics performance.

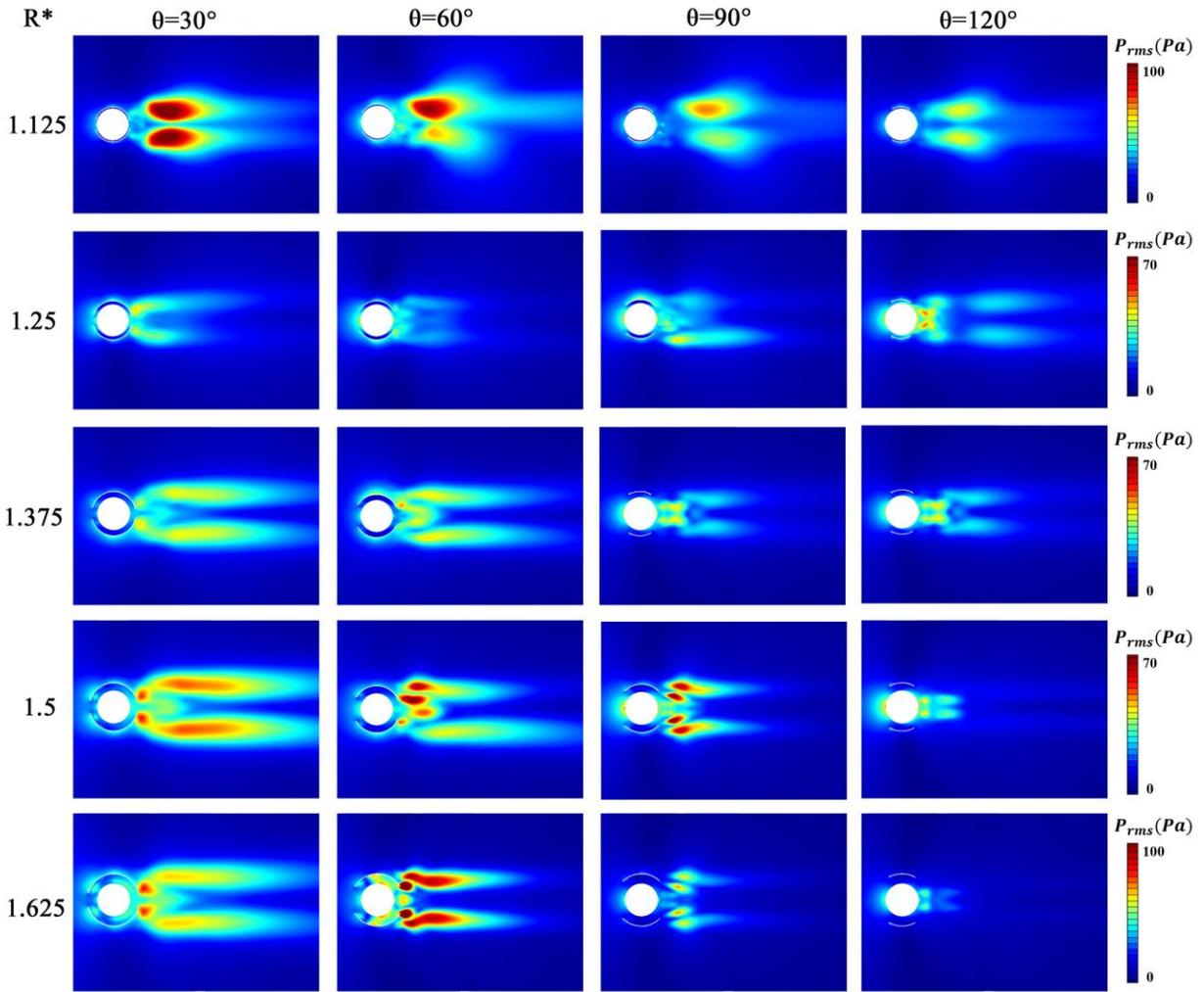

**Fig. 12.** Iso-contours of root-mean-square of static pressure ($P_{rms}$) around the equipped circular cylinder utilizing annular plates as a passive control method, in majority cases, the intensity of pressure gradient in the vicinity of structural boundaries which is main noise source in very low level.

Figure 13 highlights some of the best SPL results, corresponding to cases with lower pressure gradients throughout the domain. It is important to note that, in our simulation, the cylinder is the sole noise source. Therefore, when selecting the best SPL cases, lower pressure intensity near the annular plate boundaries is also considered. As shown in Fig. 13, a significant reduction in noise generation is observed. In the case of $D^* = 1.25, \theta = 60°$, a dramatic noise reduction of 39 dB is achieved compared to the bare cylinder, aligning with our noise reduction objectives. Moreover, this configuration exhibits remarkable aerodynamic characteristics. The Strouhal number and mean drag coefficient are reduced by 86.6% and 50.62%, respectively. Therefore, from both aerodynamic and aeroacoustics perspectives, this configuration can be considered the optimal case.

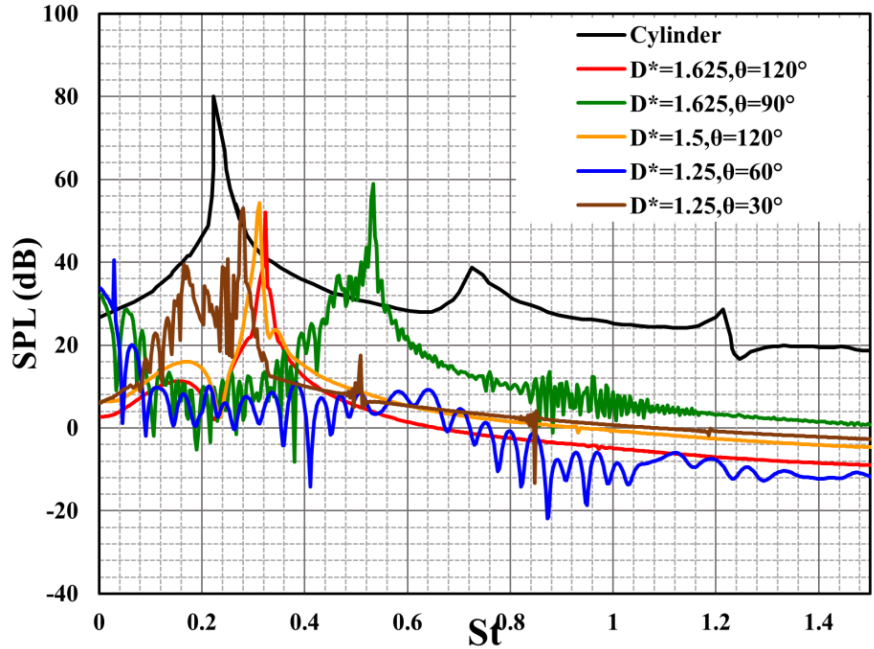

**Fig. 13.** Sound pressure level (SPL) again Strouhal number according to noise generating of circular cylinder in origin and receivers that placed in (0, 86.25D).

## 4. Conclusions

Aerodynamic noise from wind turbines is a dominant source of noise pollution in wind farms, impacting both human and animal life. Therefore, it is crucial to develop eco-friendly engineering designs to mitigate this issue. Circular structures are widely used in various engineering applications, particularly in the wind industry. This study investigates the aeroacoustics and aerodynamic performance of a two-dimensional circular cylinder using the URANS method and the FW-H acoustic analogy. To achieve noise reduction and wake control, two thin annular plates are introduced in the circumferential space of the cylinder, forming a passive control strategy. A comprehensive parametric study is conducted based on the normalized diameter ($D^*$) and the open angle ($\theta$) of the annular plates to determine the optimal configuration. The drag coefficient, Strouhal number, and sound pressure level (SPL) are evaluated for different control configurations. The results demonstrate that applying this passive control method significantly reduces pressure fluctuations and aerodynamic forces. A strong correlation between drag reduction and the inclusion of annular plates is confirmed through the investigation of the mean pressure coefficient. Specifically, a maximum drag reduction of 89.46% is observed for $D^* = 1.25$, $\theta = 30°$, compared to the bare cylinder. From both aerodynamic and aeroacoustics perspectives, the optimal configuration is $D^* = 1.25$, $\theta = 60°$, where noise generation is reduced by 39 dB. In this case, the Strouhal number and mean drag coefficient decrease by 86.6% and 50.62%, respectively. Due to its advantages such as ease of installation, simplicity, and effectiveness in noise reduction the proposed passive control method has potential applications in wind turbine design. Furthermore, its strong control capability at high Reynolds numbers, which are more relevant in practical engineering scenarios, broadens the scope of its applicability.

86. Doolan, C.J., *Computational bluff body aerodynamic noise prediction using a statistical approach.* Applied Acoustics, 2010. **71**(12): p. 1194-1203.
87. Zhu, H., W. Liu, and T. Zhou, *Direct numerical simulation of the wake adjustment and hydrodynamic characteristics of a circular cylinder symmetrically attached with fin-shaped strips.* Ocean Engineering, 2020. **195**: p. 106756.
88. Ma, R., et al., *Acoustic analysis of a forced-oscillating cylinder in flow using a hybrid method.* Aerospace Science and Technology, 2020. **106**: p. 106137.
89. Benard, N., et al., *Control of diffuser jet flow: turbulent kinetic energy and jet spreading enhancements assisted by a non-thermal plasma discharge.* Experiments in fluids, 2008. **45**(2): p. 333-355.
90. Chen, W.-L., H. Li, and H. Hu, *An experimental study on a suction flow control method to reduce the unsteadiness of the wind loads acting on a circular cylinder.* Experiments in fluids, 2014. **55**(4): p. 1-20.
91. Curle, N., *The influence of solid boundaries upon aerodynamic sound.* Proceedings of the Royal Society of London. Series A. Mathematical and Physical Sciences, 1955. **231**(1187): p. 505-514.
24